\documentclass[12pt,a4paper]{article}
\begin{document}
\large

\newpage
\begin{center}
{\bf ON THE GRAVITATIONAL FIELD OF THE ELECTRIC CHARGE}
\end{center}
\vspace{1cm}
\begin{center}
{\bf Rasulkhozha S. Sharafiddinov}
\end{center}
\vspace{1cm}
\begin{center}
{\bf Institute of Nuclear Physics, Uzbekistan Academy of Sciences,
Tashkent, 702132 Ulugbek, Uzbekistan}
\end{center}
\vspace{1cm}

The idea has been speaked out that any of Dirac and Pauli form factors of
leptonic current includes in self both normal and anomalous components. From
this point of view, with the use of cross sections of elastic scattering of
unpolarized and longitudinal polarized electrons and their neutrinos by
spinless nuclei the dependence of independent parts of charge and magnetic
moment is established. Some considerations of a connection between the mass
of a particle and its electromagnetic nature are listed which can explain
also the appearance of gravitational field of a Coulomb interaction. All
they state that a massive four - component neutrino similarly to the electron
must have the normal as well as the anomalous electric charge.

\newpage
Owing to the Dirac nature of mass, the neutrino can interact with all gauge
bosons which plays an important part in establishing the physical picture of
massive fermions. Here the processes on the nuclear targets are particularly
interesting because they give the possibility to investigate of compound
structure of an incoming particle \cite{1,2} as well as of the unified
system of hadrons \cite{3} themselves.

The interaction of light leptons $(l=e,\nu_{e})$ with the virtual photon
strongly depends on the two scalar functions $F_{1l}(q^{2})$ and
$F_{2l}(q^{2})$ are the Dirac and Pauli form factors of vector current.
Of them $F_{1l}(0)$ defines the electric charge: $F_{1l}(0)=e_{l}.$ It
appears that $F_{2l}(0)$ gives only the anomalous magnetic moment \cite{4},
and a particle full magnetic moment will have \cite{5} the estimate \cite{6}
of $\mu_{l}^{full}=(F_{1l}(0)/2m_{l})+F_{2l}(0).$

We assume that each of existing types of magnetic moments must arise
as a consequence of the availability of a kind of charge. From this point
of view, the functions $F_{il}(q^{2})$ may be written as
\begin{equation}
F_{il}(q^{2})=f_{il}(0)+A_{il}(\vec{q^{2}})+...,
\label{1}
\end{equation}
where $f_{il}(0)$ are the normal charge and moment:
$$f_{1l}(0)=e_{l}^{norm}, \, \, \, \,
f_{2l}(0)=\mu_{l}^{norm}.$$

The terms $A_{il}(\vec{q^{2}})$ characterize the anomalous behavior
of form factors and in the low energy limit $(\vec{q^{2}}=0)$ have
the size
$$A_{1l}(0)=e_{l}^{anom}, \, \, \, \, A_{2l}(0)=\mu_{l}^{anom}.$$

Such a sight to the interaction nature can also explain the fact that
$F_{1l}(0)$ and $F_{2l}(0)$ give the full static values of a Dirac
particle electric charge and magnetic moment:
\begin{equation}
F_{1l}(0)=e_{l}^{full}=e_{l}^{norm}+e_{l}^{anom}+...,
\label{2}
\end{equation}
\begin{equation}
F_{2l}(0)=\mu_{l}^{full}=\mu_{l}^{norm}+\mu_{l}^{anom}+....
\label{3}
\end{equation}

According to the hypothesis of field mass based on the classical theory
of an extensive electron \cite{7}, a particle all the mass is strictly
electromagnetic. Therefore, from point of view of compound structure
of form factors, it should be expected that any non - zero component of
the electric charge implies the existence of a kind of mass. In other words,
the lepton mass $m_{l}$ contains as well as the normal $m_{l}^{norm}$ and
anomalous $m_{l}^{anom}$ parts:
\begin{equation}
m_{l}=m_{l}^{E}=m_{l}^{norm}+m_{l}^{anom}+....
\label{4}
\end{equation}

The purpose of our article is to refine the vector picture of
massive neutrinos investigating the elastic scattering of unpolarized
and longitudinal polarized electrons and their neutrinos on a spinless
nucleus going through the interactions of charge and magnetic moment of
light leptons with field of emission of virtual photons.

The correspondence principle states that in the case of one - photon
exchange only the currents independent components are responsible for
this process. Insofar as $A_{il}(\vec{q^{2}})$ are concerned, the
inclusion of such terms in the discussion will require the account
of the second Born approximation.

At these conditions, the scattering amplitude in the normal limit
has the following structure:
\begin{equation}
M^{em}_{fi}=\frac{4\pi\alpha}{q^{2}}\overline{u}(p',s')
[\gamma_{\mu}f_{1l}(0)-i\sigma_{\mu\lambda}q_{\lambda}f_{2l}(0)]
u(p,s)<f|J^{\gamma}_{\mu}(q)|i>.
\label{5}
\end{equation}
Here $l=e_{L,R}$ or $\nu=\nu_{L,R}=\nu_{eL,R}, \, \, \, \,
\sigma_{\mu\lambda}=[\gamma_{\mu},\gamma_{\lambda}]/2, \, \, \, \,
q=p-p', \, \, \, \, p (p')$ and $s (s')$ are the four - momentum (helicities)
of a particle before and after the interaction, $J_{\mu}^{\gamma}$ is the
nuclear vector current \cite{8}.

The cross sections of the studied processes on the basis of (\ref{5}) one
can present as
$$\frac{d\sigma_{em}^{V_l}(\theta_{l},s,s')}{d\Omega}=
\frac{1}{2}\sigma^{l}_{o}(1-\eta^{2}_{l})^{-1}\{(1+ss')f_{1l}^{2}+$$
\begin{equation}
+\eta^{2}_{l}(1-ss')[f_{1l}^{2}+
4(m_{l}^{norm})^{2}(1-\eta^{-2}_{l})^{2}f_{2l}^{2}]
tg^{2}\frac{\theta_{l}}{2}\} F^{2}_{E}(q^{2}),
\label{6}
\end{equation}
where we must have in view of that
$$\sigma_{o}^{l}=
\frac{\alpha^{2}cos^{2}\frac{\theta_{l}}{2}}{4E^{2}_{l}(1-\eta^{2}_{l})
sin^{4}\frac{\theta_{l}}{2}}, \, \, \, \, \eta_{l}=
\frac{m_{l}^{norm}}{E_{l}},$$
$$E_{l}=\sqrt{p^{2}+(m_{l}^{norm})^{2}}, \, \, \, \, F_{E}(q^{2})=
ZF_{c}(q^{2}).$$
Here $\theta_{l}$ is the scattering angle, $F_{E}(q^{2})$ is the nucleus
electric $(F_{E}(0)=Z)$ form factor, $E_{l}$ and $m_{l}^{norm}$ are the
fermion normal mass and energy, and the presence of $V_{l}$ implies the
absence of the current axial - vector part.

Taking into account that (\ref{6}) describes such interactions as
conserving $(s' = s)$ and nonconserving $(s'=-s)$ the left $(s=-1)$ - and
right (s=+1) - handed neutrinos helicities, we may rewrite its in the form
\begin{equation}
d\sigma_{em}^{V_{l}}(\theta_{l},s)=
d\sigma_{em}^{V_{l}}(\theta_{l},f_{1l},s)+
d\sigma_{em}^{V_{l}}(\theta_{l},f_{2l},s),
\label{7}
\end{equation}
$$\frac{d\sigma_{em}^{V_{l}}(\theta_{l},f_{1l},s)}{d\Omega}=
\frac{d\sigma_{em}^{V_{l}}(\theta_{l},f_{1l},s'=s)}{d\Omega}+
\frac{d\sigma_{em}^{V_{l}}(\theta_{l},f_{1l},s'=-s)}{d\Omega}=$$
\begin{equation}
=\sigma^{l}_{o}(1-\eta^{2}_{l})^{-1}(1+\eta_{l}^{2}tg^{2}
\frac{\theta_{l}}{2})f_{1l}^{2}F_{E}^{2}(q^{2}),
\label{8}
\end{equation}
$$\frac{d\sigma_{em}^{V_{l}}(\theta_{l},f_{2l},s)}{d\Omega}=
\frac{d\sigma_{em}^{V_{l}}(\theta_{l},f_{2l},s'=-s)}{d\Omega}=$$
\begin{equation}
=4(m_{l}^{norm})^{2}\sigma^{l}_{o}(1-\eta^{2}_{l})\eta^{-2}_{l}f_{2l}^{2}
F_{E}^{2}(q^{2})tg^{2}\frac{\theta_{l}}{2}.
\label{9}
\end{equation}

As seen from (\ref{8}) and (\ref{9}), in the case of massive neutrinos,
the charge is responsible for the scattering either with or without flip
of spin. Insofar as the magnetic moment is concerned, it changes a particle
helicity.

Averaging over $s$ and summing over $s',$  the cross section (\ref{6}) one
can reduce to the following:
\begin{equation}
d\sigma_{em}^{V_{l}}(\theta_{l})=
d\sigma_{em}^{V_{l}}(\theta_{l},f_{1l})+
d\sigma_{em}^{V_{l}}(\theta_{l},f_{2l}),
\label{10}
\end{equation}
\begin{equation}
\frac{d\sigma_{em}^{V_{l}}(\theta_{l},f_{1l})}{d\Omega}=
\sigma^{l}_{o}(1-\eta^{2}_{l})^{-1}
(1+\eta^{2}_{l}tg^{2}\frac{\theta_{l}}{2})f_{1l}^{2}F_{E}^{2}(q^{2}),
\label{11}
\end{equation}
\begin{equation}
\frac{d\sigma_{em}^{V_{l}}(\theta_{l},f_{2l})}{d\Omega}=
4(m_{l}^{norm})^{2}\sigma^{l}_{o}(1-\eta^{2}_{l})\eta^{-2}_{l}f_{2l}^{2}
F_{E}^{2}(q^2)tg^{2}\frac{\theta_{l}}{2}.
\label{12}
\end{equation}

Comparison of (\ref{7}) and (\ref{10}) shows clearly that
\begin{equation}
\frac{d\sigma_{em}^{V_{l}}(\theta_{l},s)}
{d\sigma_{em}^{V_{l}}(\theta_{l})}=1,
\label{13}
\end{equation}
which leads to the system of the six the most diverse equations.

For our purposes it is desirable to choose only the two of them:
\begin{equation}
\frac{d\sigma_{em}^{V_{l}}(\theta_{l},f_{2l},s)}
{d\sigma_{em}^{V_{l}}(\theta_{l},f_{1l},s)}=1,
\label{14}
\end{equation}
\begin{equation}
\frac{d\sigma_{em}^{V_{l}}(\theta_{l},f_{2l})}
{d\sigma_{em}^{V_{l}}(\theta_{l},f_{1l})}=1.
\label{15}
\end{equation}

The basis for such a statement is that both numerator and denominator
in (\ref{14}) and (\ref{15}) there exist simultaneously.

According to our description, the neutrino spin flip proceeding in the
Coulomb field is observed because of an intimate connection between the
mass of the neutrino and its electromagnetic nature \cite{1}. At the same
time a question of the structure and properties of mass responsible
for conservation as well as for change of a particle helicity requires
the special investigation.

But here we can use of (\ref{15}), so as it gives the possibility to
establish the corresponding picture of the interaction regardless of spin
polarization. Indeed, inserting (\ref{11}) and (\ref{12}) in (\ref{15}),
it is easy to observe the following dependence of form factors
\begin{equation}
4(m_{l}^{norm})^{2}\frac{f_{2l}^{2}(0)}{f_{1l}^{2}(0)}=
\frac{\eta_{l}^{2}(1+\eta_{l}^{2}tg^{2}\frac{\theta_{l}}{2})}
{(1-\eta_{l}^{2})^{2}tg^{2}\frac{\theta_{l}}{2}}.
\label{16}
\end{equation}

Furthermore, if neutrinos are of low energies
$(E_{l}\rightarrow m_{l}^{norm})$ then
$$lim_{\eta_{l}\rightarrow 1}
\frac{\eta_{l}^{2}(1+\eta_{l}^{2}tg^{2}\frac{\theta_{l}}{2})}
{(1-\eta_{l}^{2})^{2}tg^{2}\frac{\theta_{l}}{2}}=1,$$
and the relation (\ref{16}) is reduced to the form
\begin{equation}
2m_{l}^{norm}\frac{f_{2l}(0)}{f_{1l}(0)}=\pm 1.
\label{17}
\end{equation}

Using (\ref{17}) for the electron with charge $f_{1e}(0)=-e_{e}^{norm},$
we find that
\begin{equation}
f_{2e}(0)=\mp\frac{e_{e}^{norm}}{2m_{e}^{norm}}.
\label{18}
\end{equation}

In the framework of the standard model of electroweak interactions,
the neutrino anomalous magnetic moment arises as a consequence of the
availability of the neutrino non - zero rest mass \cite{9}. It is of course
not excluded that if start from this connection, the function $A_{2\nu}(0)$
must have the form
\begin{equation} A_{2\nu}(0)=\mu_{\nu}^{anom}=
\frac{3eG_{F}m_{\nu}^{anom}}{8\pi^{2}\sqrt{2}}, \, \, \, \, e=|e_{e}^{norm}|.
\label{19}
\end{equation}

Turning to (\ref{4}) and (\ref{19}), for the normal part of the neutrino
magnetic moment, we get
\begin{equation}
f_{2\nu}(0)=\mu_{\nu}^{norm}=
\frac{3eG_{F}m_{\nu}^{norm}}{8\pi^{2}\sqrt{2}},
\label{20}
\end{equation}
and therefore, the solution of equation (\ref{17}) may behave as
\begin{equation}
f_{1\nu}(0)=e_{\nu}^{norm}=
-\frac{3eG_{F}(m_{\nu}^{norm})^{2}}{4\pi^{2}\sqrt{2}}.
\label{21}
\end{equation}

Insofar as the neutrino anomalous electric charge is concerned, with the aid
of (\ref{4}) one can find from (\ref{21}) that
\begin{equation}
A_{1\nu}(0)=e_{\nu}^{anom}=
-\frac{3eG_{F}(m_{\nu}^{anom})^{2}}{4\pi^{2}\sqrt{2}}.
\label{22}
\end{equation}

Insertion of (\ref{19}) - (\ref{22}) in (\ref{2}) ¨ (\ref{3}) defines
the full static sizes of the neutrino electric charge and magnetic moment.
Their general structure at the account of (\ref{4}) has the form
\begin{equation}
e_{\nu}=e_{\nu}^{full}=-\frac{3eG_{F}m_{\nu}^{2}}{4\pi^{2}\sqrt{2}},
\label{23}
\end{equation}
\begin{equation}
\mu_{\nu}=\mu_{\nu}^{full}=\frac{3eG_{F}m_{\nu}}{8\pi^{2}\sqrt{2}}.
\label{24}
\end{equation}

The availability of a connection between $\mu_{\nu}$ and $m_{\nu}$
can also confirm the existence of gravitational field of magnetic
moment \cite{10}. Of course, such a mass dependence and testifies
of that the electric mass and charge of a particle correspond to two
form of the unified regularity of its physical nature \cite{11}. Owing
to this, becomes possible move from the Newton law of gravity to the
Coulomb law and vice versa.

These forces for two neutrinos can be expressed in the form
\begin{equation}
F_{N}=G_{N}\frac{m_{\nu}^{2}}{R^{2}}, \, \, \, \,
F_{C}=\frac{1}{4\pi\epsilon_{0}}\frac{e_{\nu}^{2}}{R^{2}},
\label{25}
\end{equation}
where $G_{N}$ is the constant of gravitational field.

Comparing (\ref{23}) with (\ref{25}), taking \cite{12}
$G_{N}=6.70916\cdot 10^{-39}\ {\rm GeV^{-2}}$ and having in view
of the inequality $F_{C}>F_{N},$ we are led to the implication that
\begin{equation}
m_{\nu}>\frac{4\pi^{2}\sqrt{2}}{3G_{F}}
\left(\frac{G_{N}}{\alpha}\right)^{1/2}=1.53\cdot 10^{-3}\ {\rm eV},
\label{26}
\end{equation}
\begin{equation}
e_{\nu}>\frac{4\pi^{2}\sqrt{2}}{3G_{F}}
\left(\frac{G_{N}}{\alpha}\right)e=1.46\cdot 10^{-30}\ {\rm e}.
\label{27}
\end{equation}

Finally, insofar as the steadity of the distribution of a massive
particle electric charge, gauge invariance and charge conservation
and quantization laws are concerned, all they together with some
aspects of compound structure of the neutrino mass will be presented
in the separate work.

\end{document}